\documentstyle[aps,multicol,epsf,psfig]{revtex}
\newcommand{\ket}[1]{\left | #1 \right \rangle}
\newcommand{\eq}{\begin{equation}}
\newcommand{\en}{\end{equation}}

\begin{document}

\title{Absorption-free discrimination between
    Semi-Transparent Objects}
\author{Graeme Mitchison$^{1}$ and Serge Massar$^{2}$}
\address{$^{1}$M.R.C. Laboratory of Molecular Biology, Hills Road,
Cambridge CB2 2QH, UK.\\
 $^{2}$Service de Physique Th\'eorique,
Universit\'e Libre de Bruxelles, C.P. 225, Bvd. du Triomphe, B1050
Bruxelles, Belgium.}

\maketitle

\begin{abstract}
Absorption-free (also known as ``interaction-free'') measurement aims
to detect the presence of an opaque object using a test particle
without that particle being absorbed by the object.  Here we consider
semi-transparent objects which have an amplitude $\alpha$ of
transmitting a particle while leaving the state of the object
unchanged and an amplitude $\beta$ of absorbing the particle.  The
task is to devise a protocol that can decide which of two known
transmission amplitudes is present while ensuring that no particle
interacts with the object.  We show that the probabilities of being
able to achieve this are limited by an inequality. This inequality
implies that absorption free distinction between complete transparency
and any partial transparency is always possible with probabilities
approaching 1, but that two partial transparencies can only be
distinguished with probabilities less than 1.
\end{abstract}

\vspace{-1.1cm}
\begin{multicols}{2}

In ``interaction-free'' measurement the task is to decide, using a
test particle, whether an opaque object is present or absent while
ensuring that the test particle is not absorbed by the object. Many
methods for achieving this have been devised
\cite{Renninger60,Dicke81,ElitzurVaidman93,KwiatEtal95,KwiatEtal99,PaulPavicic97,TsegayeEtal98}.
The essential idea behind all of them is that the measurement picks a
set of histories in none of which an interaction between object and
test particle takes place, so no absorption occurs; other histories in
the protocol will involve interactions, and in this sense the term
``absorption-free'' may be preferred to the more commonly-used term
``interaction-free''. We abbreviate absorption-free measurement to AFM
henceforth.

In standard AFM, the object is considered to be either completely
opaque or completely transparent (absent).  One can also consider
semi-transparent objects, for which there is an amplitude $\alpha$ of
the particle passing through the object while leaving the state of the
object unchanged and an amplitude $\beta$ for the particle to interact
with the object and hence be absorbed, leaving the object in an
``interacted'' state (in Elitzur and Vaidman's proposal this
is the exploded state of the bomb). One can then ask whether one can
infer the transmission amplitude of the object while ensuring that the
object never reaches the ``interacted'' state. This is the problem we
consider here, in the case where there are two known transmission
amplitudes that have to be distinguished.

This problem is of obvious practical interest. Indeed there are
situations where one wants to determine the nature of an object but
where radiation will damage the object, for instance when imaging a
biological specimen in the ultraviolet. In these cases one wants to
minimize the amount of radiation absorbed by the object. Standard AFM
shows that if the object has only two possible states, completely
transparent or completely opaque, then it is possible to determine the
state without any photon being absorbed. However most objects will be
semi-transparent. Here we address this more general case.

In \cite{MitchisonJozsa99}, a general framework for counterfactual
quantum events was proposed, which includes AFM. Two variables
$\ket{p}$ and $\ket{q}$ are distinguished in the total state
space. The first variable $\ket{p}$ defines the state of the particle
and its position within the apparatus used for AFM, and we assume
there is a particular subset of values, $\cal I$, for which
interactions between the particle and apparatus can occur leading to
absorption. The second variable $\ket{q}$, which we call the {\em
interaction variable}, takes the value $\ket{1}$ if absorption occurs
and $\ket{0}$ if not. It may have additional values, but they play no
role in the following discussion. 

Any protocol for AFM can be divided into a series of steps. In some of
these steps an interaction can potentially occur; we call these
I-steps. An I-step has two parts. The first is a unitary
transformation given by
\begin{equation}
\begin{array}{ll}
\ket{p0} \rightarrow \alpha \ket{p0}+\beta \ket{p1} &p \in \cal I \\
\ket{p0} \rightarrow \ket{p0}  &p \notin \cal I
\label{eq:amp}
\end{array}
\end{equation}
where $\alpha$, $\beta$ are complex numbers satisfying
$|\alpha|^2+|\beta|^2=1$. The second part is a measurement of the
interaction variable in the basis $\ket{0}$, $\ket{1}$. The unitary
transformation is not fully defined by (\ref{eq:amp}), but we do not
need to specify its action on terms like $\ket{p1}$ since we are
concerned with histories on which no interaction occurs, and a protocol
can be assumed to halt when measurement of the interaction variable
yields $\ket{1}$\footnote{In the case of photons, we can also let many
photons pass through the object together. The I-step then takes the
form $|n0\rangle \to \alpha^n |n0\rangle +|interacted\ states\rangle$
(if the photons belong to $\cal I$), and after the measurement of the
interaction variable, the state becomes $\alpha^n |n0\rangle$ (if no
interaction occurs). This is identical, up to a unitary
transformation, to the state obtained if n particles pass successively
through the object and none interact with the object. This remark
shows that restricting to particles passing one by one, as
in (\ref{eq:amp}), does not make the analysis less general.  }.

A protocol for AFM starts from a specified initial state. It is
allowed to undergo a unitary transformation between successive
I-steps, this transformation leaving the interaction variable
unchanged. At the end of the protocol the variable $\ket{p}$ is
measured. A protocol with measurements of $\ket{p}$ before the end can
be converted to the form we specify by entangling the measured
variables with extra variables and postponing their measurement till
the end \cite{MitchisonJozsa99}.

In all the protocols, there are two measurement outcomes, $M_1$ and
$M_2$ say, the first of which indicates that the object was absent,
while the second indicates that the object was present and also that
no absorption occurred. There will also be other outcomes, for
instance that an absorption occurred. We denote the probability of
$M_i$ by $P(ident|i)$, i.e. the probability of identifying, without
the particle being absorbed, whether the object is present ($i=2$) or
absent ($i=1$). The probabilities $P(ident|i)$ give an indication of
the efficiency of the protocol. In Elitzur and Vaidman's original
proposal \cite{ElitzurVaidman93}, one has\footnote{In Elitzur and
Vaidman's original proposal one can never be certain that the object
is absent, hence $P(ident|1)=0$, and the probability of learning that
the bomb is present without it exploding is $1/4$.} $P(ident|2)=1/4$ and 
 $P(ident|1)=0$. In many recent protocols, $P(ident|2)=1$ and
$P(ident|1)$ tends to 1.

Figure 1 shows two types of AFM protocol. The quantum Zeno type
\cite{KwiatEtal95,KwiatEtal99} is an elaboration of Elitzur and
Vaidman's original proposal \cite{ElitzurVaidman93}.  We have adapted
it so that it can distinguish between no object ($\alpha_1=1$) and an
object of transmission amplitude $0\leq |\alpha_2| <1$. We take the
first qubit of $|pq\rangle$ to correspond to polarization, and the
initial state is a vertically polarized photon, denoted
$|v0\rangle$. The AFM consists of repeated passages through a
polarization rotator, a Mach-Zender interferometer, and a second
polarization rotator. After the first rotation the state becomes $\cos
\theta |v0\rangle + \sin \theta |h0\rangle$. After passing through the
polarization beam splitter, the horizontally polarised component
$\ket{h0}$ goes along the lower path, that may contain the object,
whereas $\ket{v0}$ takes the object-free upward-going path. In this
case, therefore, $\cal I$ is the single value $p=h$.  Applying
(\ref{eq:amp}), the I-step gives the un-normalised state
$\ket{\psi_i}=\cos \theta\ket{v0}+\alpha_i \sin\theta\ket{h0}$.  The
second polarization beam splitter then recombines the two polarizations
into one beam.  If $\alpha_2$ is real and positive, then the state in
case $i=2$ can be rewritten as $\ket{\psi_2}=\gamma \left( \cos
\theta'\ket{v0}+\sin\theta'\ket{h0}\right)$ where $\cos \theta' = \cos
\theta / \sqrt{\cos^2\theta + \alpha_2^2 \sin^2\theta}$ (note that
$\theta' \leq \theta$).  The final step is a rotation by the angle
$-\theta'$. This brings the state to $\cos(\theta-\theta')|v0\rangle +
\sin(\theta-\theta')\ket{h0}$ (no object present) or $\gamma
|v0\rangle$ (object present).  We then iterate this procedure $N$
times, choosing $N$ such that $N ( \theta-\theta') = \pi /2$. This
brings the state to $|h0\rangle$ (no object present) or $\gamma^N
|v0\rangle$ (object present). Since these states are orthogonal
$P(ident|1)=1$ and $P(ident|2)=\gamma^{2N}$. For large $N$ (small
$\theta$), $\gamma \simeq 1-{(1+\alpha_2) \pi^2 \over (1-\alpha_2) 4
N^2}$ and $P(ident|2)\to 1$.

\begin{figure}
%\centerline{\psfig{figure=eps/ifm2.eps,width=0.9\textwidth}}
\centerline {\psfig{width=7.0cm,file=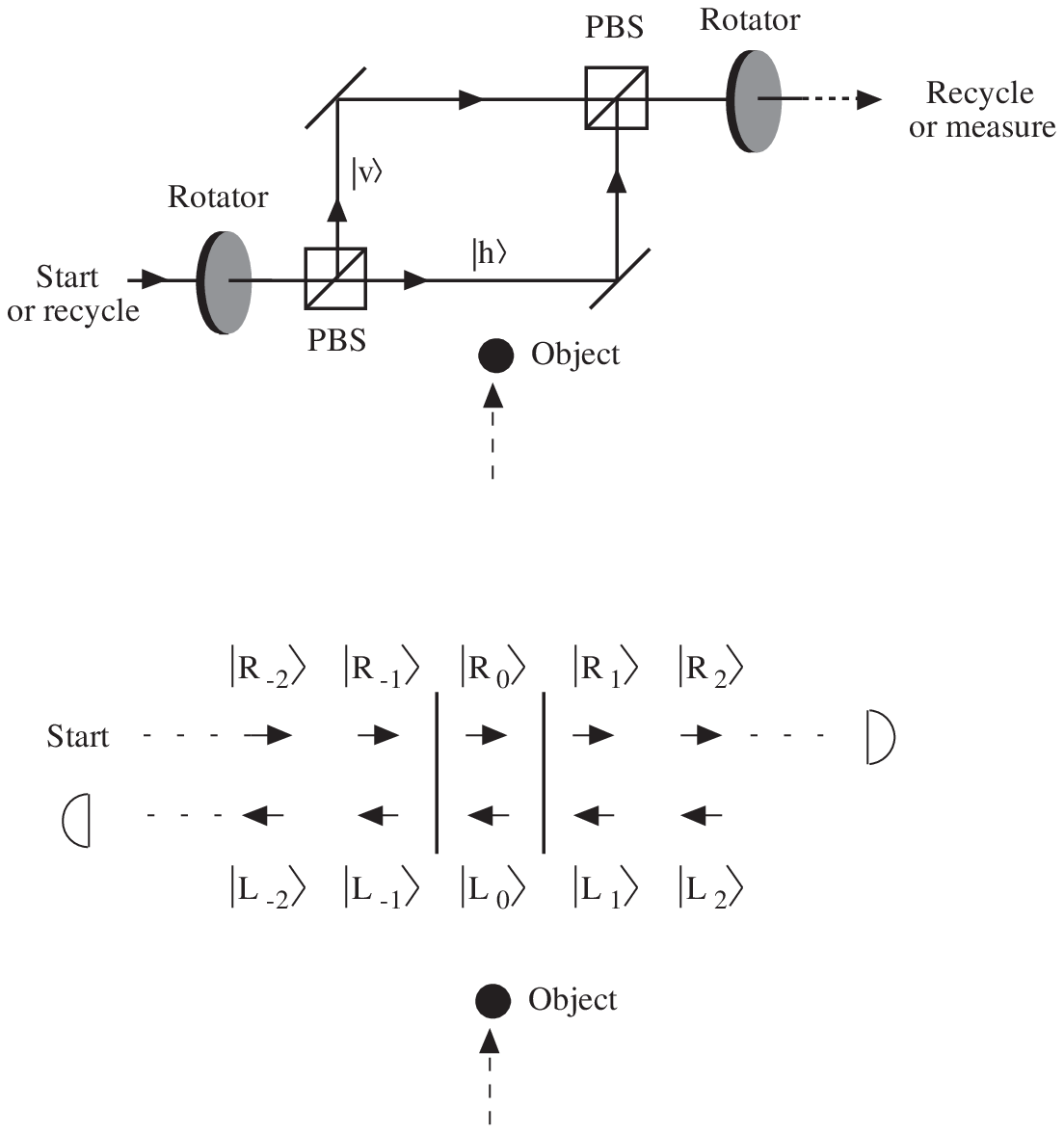}}
\begin{narrowtext}
\caption{ Two absorption-free measurement protocols. In the quantum
Zeno type (above) the photon is passed through a rotator, followed by
two polarising beam-splitters (PBS) and a final polarisation rotator,
and is then recycled a certain number of times. In our protocol for
$\alpha_1=1$, $\alpha_2$ real, the first rotation is by an angle
$\theta$ and the second is by an angle $(\theta-\theta')$ (see
text). In the Fabry-Perot type (below), the plane wave representing
the photon is decomposed into a superposition of wave packets. The
wave function for a right-going pulse at position $i$ is denoted by
$\ket{R_i}$ and for a left-going pulse by $\ket{L_i}$. If an object is
inserted in the cavity (between the two vertical lines) the photon may
be reflected to the left. If there is no object, it will be
transmitted to the right.
\label{fig:example2}}
\end{narrowtext}
\end{figure}

If $\alpha_2 = |\alpha_2| e^{i\phi_2}$ has a non zero phase (the phase
is defined by the convention that $\alpha_1=1$) then after recombining
the two beams the state is $\ket{\psi_2}=\gamma \left( \cos
\theta'\ket{v0}+\sin\theta'e^{i\phi_2} \ket{h0}\right)$. The final
rotation is chosen so as to take this state to the state $\gamma
|v0\rangle$ and to take the state if no object is present to
$\cos\omega |v0\rangle + \sin \omega |h0\rangle$ where $\theta-\theta'
\leq \omega \leq \theta+\theta'$. Iterating this procedure $N$ times,
with $N\omega = \pi/2$, and then carrying out a measurement of
polarisation, realizes an AFM.

An alternative type of AFM protocol uses the concept of a monolithic
total-internal-reflection resonator \cite{PaulPavicic97}, or a
Fabry-Perot (F-P) interferometer \cite{TsegayeEtal98}.  In the case of
the F-P there is a photon of momentum $k$ incoming from the left, and
one measures whether the photon is reflected or transmitted. To see
how the F-P fits into our framework for AFM, we make the dynamics
discrete to correspond to the steps in a protocol. Define a lattice of
spacing $d$ (the spacing of the mirrors, see Figure 1(b)) and let each
step in the protocol correspond to a time $t=d/c$. The state
$|R_n\rangle$ corresponds to a segment of right-moving plane wave over
the spatial range $[nd,(n+1)d]$ for time interval $t$, and
$|L_n\rangle$ to the corresponding segment of left-moving plane
wave. Over time $t$, $|R_n\rangle$ evolves into $|R_{n+1}\rangle$ and
$|L_n\rangle$ into $|L_{n-1}\rangle$, except in the vicinity of the
mirrors, where we have $\ket{R_0} \rightarrow c\ket{L_0}+is\ket{R_1}$,
$\ket{R_{-1}} \rightarrow c\ket{L_{-1}}+is\ket{R_0}$, $\ket{L_1}
\rightarrow c\ket{R_1}+is\ket{L_0}$, and $\ket{L_0} \rightarrow
c\ket{R_0}+is\ket{L_{-1}}$, $c$ and $is$ being reflection and
transmission coefficients, respectively.

We have treated the mirrors as dispersionless, which is a mathematical
convenience to restrict ourselves to the Fourier component of the
incoming plane wave. We have also taken $d$ such that $e^{idk}=1$
(where $k$ is the wave number of the plane wave) so
that no phase is accumulated between the mirrors.

If an object of transparency $\alpha$ is inserted between the two
mirrors, the discretised dynamics, conditional on no photon being
absorbed, becomes $\ket{R_0} \rightarrow \alpha\left(
c\ket{L_0}+is\ket{R_1}\right)$, $\ket{R_{-1}} \rightarrow
c\ket{L_{-1}}+is\ket{R_0}$, $\ket{L_1} \rightarrow
c\ket{R_1}+is\ket{L_0}$, and $\ket{L_0} \rightarrow\alpha\left(
c\ket{R_0}+is\ket{L_{-1}}\right)$. Thus ${\cal I}=\{L_0, R_0\}$,
since interaction can only occur within the apparatus. 

The initial state is $e^{ikx}=\sum^{n=-1}_{-\infty} \ket{R_n}$. After
many time steps, one settles into a steady state regime and the state
outgoing to the left $f_L$ is the sum of the pulses reflected once by
the left mirror and those reflected $2m-1$ times inside the instrument
and traversing the object $2m$ times, for $m=1, 2, \ldots$:
\begin{eqnarray*}
f_L&=& \sum^{n=-1}_{-\infty}\ket{L_n} \left(c+s^2\sum^{\infty}_{m=1}c^{2m-1}\alpha^{2m} \right)\\
&=&\sum^{n=-1}_{-\infty}\ket{L_n}\frac{c(1-\alpha^2)}{1-c^2\alpha^2}.
\end{eqnarray*}
As $c \to 1$, the probability $|f_L|^2=|c(1-\alpha^2)/(1-c^2\alpha^2)|^2$ of
reflection to the left tends to 1, except when $\alpha=1$ (object
absent) in which case the probability of transmission to the right is
1. Thus the F-P allows an absorption-free discrimination between the
absence of an object ($\alpha_1=1$) and the presence of an object of
transparency $\alpha_2 \ne 1$.

Now consider any protocol that falls within our general scheme, and
suppose that one must distinguish between two semi-transparent objects
with transmission amplitude $\alpha_1$, $ \alpha_2$ (which can both be
different from $1$) and interaction amplitude $\beta_1$, $\beta_2$
respectively.  We shall prove the following constraint on the
probability $P(ident|i)$ of identifying transparency $\alpha_i$
without any absorption occurring:

{\bf Theorem: } $(1-P(ident|1))(1-P(ident|2)) \ge \eta^2$, 
where
$\eta=|\beta_1\beta_2/(1-\bar\alpha_1\alpha_2)|$.

Before giving the proof, we look at some of the consequences of this
inequality. First, note that $|1-\bar\alpha_1\alpha_2|^2-
|\beta_1\beta_2|^2= |\alpha_1-\alpha_2|^2$.
Thus $\eta \le 1$, and $\eta=1$ iff $\alpha_1=\alpha_2$. This implies
that $P(ident|1)=P(ident|2)=0$ 
when $\alpha_1=\alpha_2$, which must of course be the
case, since two equal transmission amplitudes cannot be distinguished. Whenever
$\alpha_1\ne \alpha_2$, however, the theorem allows non-zero values of
$P(ident|1)$ and $P(ident|2)$.

Another special case is when one object is completely
transparent (absent), ie.  $|\alpha_1|=1$ and $\beta_1=0$. If
$\alpha_2 \ne \alpha_1$, then $\eta=0$, and the theorem permits
$P(ident|1)=P(ident|2)=1$. That this can be achieved was shown above.

The most significant aspect of this result is that when both
$|\alpha_1|$ and $ |\alpha_2|$ are different from $1$, that is neither
object is completely transparent, then $\eta$ is strictly positive.
This implies that both $P(ident|1)$ and $P(ident|2)$ must be strictly
less than $1$. Thus it is impossible to identify two semi-transparent
objects with vanishing probability that the test particle is absorbed by the
objects. This is bad news for the applications outlined above.

{\bf Proof:} The total state space can be decomposed into two
orthogonal subspaces, the first spanned by components whose first
variable $p$ satisfies $p \notin \cal I$, and the second by components
whose first variable $p$ satisfies $p \in \cal I$. Recall that a
general protocol can be written as a series of I-steps followed by
unitary transformations.  We can write the un-normalized state for
transparency $i$ at stage $k$ of the protocol immediately before the
I-step as $\ket{u_i^k}+\ket{v_i^k}$, where $\ket{u_i^k}$ lies in the
first subspace and $\ket{v_i^k}$ in the second. Immediately after the
I-step (\ref{eq:amp}) implies that the un-normalized state is
\[
\ket{\psi_i^k}=\ket{u_i^k}+\alpha_i\ket{v_i^k}.
\]
We assume that the states $\ket{\psi_i^k}$ are all un-normalized, so
$1-|\psi_i^k|^2$ is the probability of no absorption occurring up to
stage $k$ of the protocol. After the I-step there is a unitary
transformation that carries  $\ket{\psi_i^k}$ to
$\ket{u_i^{k+1}}+\ket{v_i^{k+1}}$. We define
\begin{equation}
f^k=\langle \psi_1^k | \psi_2^k \rangle,
\label{eq:fdef}
\end{equation}
whereupon unitarity implies
\begin{eqnarray*}
f^k&=& \langle u_1^{k+1}+v_1^{k+1} | u_2^{k+1}+v_2^{k+1} \rangle \\
&=& \langle u_1^{k+1}| u_2^{k+1} \rangle + \langle v_1^{k+1} | v_2^{k+1}
\rangle 
\end{eqnarray*}
(since the components $u$ and $v$ lie in orthogonal subspaces),
and (\ref{eq:fdef}) for $k+1$ implies
\[
f^{k+1}=\langle u_1^{k+1}| u_2^{k+1} \rangle + 
\bar\alpha_1\alpha_2\langle v_1^{k+1} | v_2^{k+1} \rangle.
\]
We therefore get 
\[
f^{k+1}=f^k-(1-\bar\alpha_1\alpha_2)\langle v_1^{k+1} | v_2^{k+1} \rangle,
\]
and hence
\[
f^N=1 - (1-\bar\alpha_1\alpha_2) 
\sum_{k=0}^{N-1} \langle v_1^{k+1} | v_2^{k+1} \rangle,
\]
where the $N$-th step is the last step of the protocol before the final
measurement. This implies
\[
{ | 1- f^N |^2  \over  |1-\bar\alpha_1\alpha_2|^2 } = 
| \sum_{k=0}^{N-1} \langle v_1^{k+1} | v_2^{k+1} \rangle |^2 \ .
\]
We now use the Cauchy-Schwartz inequality to obtain
\begin{equation}
{ | 1- f^N |^2  \over  |1-\bar\alpha_1\alpha_2|^2 } \leq
 \sum_{k=0}^{N-1} | v_1^{k+1} |^2  \sum_{k=0}^{N-1} |  v_2^{k+1}|^2 \ .
\label{CS}
\end{equation}
The probability that an interaction occurs during the
k'th I-step is $|\beta_i|^2 |v^k_i|^2$, and therefore 
we can 
rewrite (\ref{CS}) as
\begin{equation}
{ | 1- f^N |^2 |\beta_1|^2 |\beta_2|^2 \over
  |1-\bar\alpha_1\alpha_2|^2 } 
\leq P(interact|1) P(interact|2) 
\label{CSfinal}
\end{equation}
where $P(interact|i)=|\beta_i|^2\sum | v_i^{k+1} |^2$ is the total
probability of interaction for transparency $i$.

We now turn to the final measurement. There are three possible
outcomes. The first is that the test particle is absorbed by the
object.  The second is that the particle is not absorbed and that the
object is identified.  The third is that no absorption occurs but the
object is not identified. This occurs with probability
$P(NOTident|i)$.  Our aim is to construct a measurement setup such
that $P(ident|i)$ is as large as possible. To this end we note that an
optimal setup necessarily has $P(NOTident|i)=0$. Indeed suppose that
$P(NOTident|i) \neq 0$. Then we run the protocol once, and if we
obtain the outcome $NOTident$ we run the protocol a second time
(constructing a protocol with its measurement of $\ket{p}$ at the end
by entangling the outcomes of the first protocol with extra
qubits). This increases the probability of identifying the object from
$P(ident|i)$ to $P(ident|i)(1+ P(NOTident|i))$. This procedure can be
iterated many times to ensure that the probability of not identifying
the object is as small as we wish.

Upon taking the limit $P(NOTident|i) \to 0$ one finds 
that $ P(interact|i) \to 1 - P(ident|i)$ and that $f^N \to 0$. The
latter limit is because if the two states $\psi^N_i$ can be identified 
with certainty, their scalar product must be zero. Thus in the limit 
$P(NOTident|i) \to 0$, (\ref{CSfinal}) tends to 
the inequality of the theorem.
 If $P(NOTident|i) \neq 0$, then
$P(ident|i)$ is necessarily 
smaller than in the limiting case, and the inequality
is also obeyed.
$\Box$

This result establishes some limits on AFM of semi-transparent
objects. It also raises various questions. First, can the bound be
attained? We showed above that this is the case if one of the objects
is transparent. The following numerical procedure suggests that
the bound can be approached very closely for any real $\alpha_i$.

Consider a quantum Zeno protocol based on a polarisation degree of
freedom as described above. We denote by
$\ket{\psi^k_i}=\ket{v0}a^k_i+\ket{h0}b^k_i$ the state for
transparency $i$ at stage $k$. Suppose $\ket{\psi^{k+1}_i}$ is
obtained from $\ket{\psi^k_i}$ by a rotation of angle $\theta^k$
followed by an I-step.  Then we have $a^{k+1}_i=\cos \theta^k
a^k_i-\sin \theta^k b^k_i$ and $b^{k+1}_i=\alpha_i(\sin \theta^k
a^k_i+\cos \theta^k b^k_i)$, for $i=1,2$. Pick $\lambda$ and require
$b^k_2/b^k_1=\lambda$ for all $k$, this being the condition
for equality in the Cauchy-Schwartz step leading to equation
(\ref{CS}). This would imply
\[
\tan\theta^k=\frac{\alpha_1b^k_1\lambda-\alpha_2b^k_2}
{\alpha_2a^k_2-\alpha_1a^k_1\lambda},
\]
which can be used to generate a series of angles $\theta^k$. However,
there is a problem with starting the procedure, since the initial
state must be the same for $i=1,2$, and so the ratio $b^0_2/b^0_1$
must be 1. Yet we wish to choose $\lambda$ freely. By taking
$b^0_i=\epsilon$, $a^0_i=\sqrt{1-\epsilon^2}$, $i=1,2$, for small
$\epsilon$, we ensure that the initial terms $b^0_i$ are small and
thereafter for larger terms $b^k_i$ the ratio $b^k_2/b^k_1$ is
$\lambda$. This means that the condition for equality in
Cauchy-Schwartz comes very close to being satisfied.

Simulations show that a simple search always comes up with a value of
$\lambda$ that makes the $\ket{\psi^N_i}$'s very close to orthogonal
after some number of steps $N$. One can therefore make a final
measurement at the $N$-th step using a POVM, in which the components
yielding the AFM outcomes $M_1$ and $M_2$ are very close to the
$\ket{\psi^N_i}$'s. By taking $\epsilon$ small enough one can make the
approach to equality of $(1-P(ident|1))(1-P(ident|2))$ and $\eta^2$ as
near as one likes for any $\alpha$'s (eg see Figure 2). It would be
interesting to prove analytically that this must be so, and also to
extend it to complex amplitudes.

\begin{figure}
\centerline{\psfig{figure=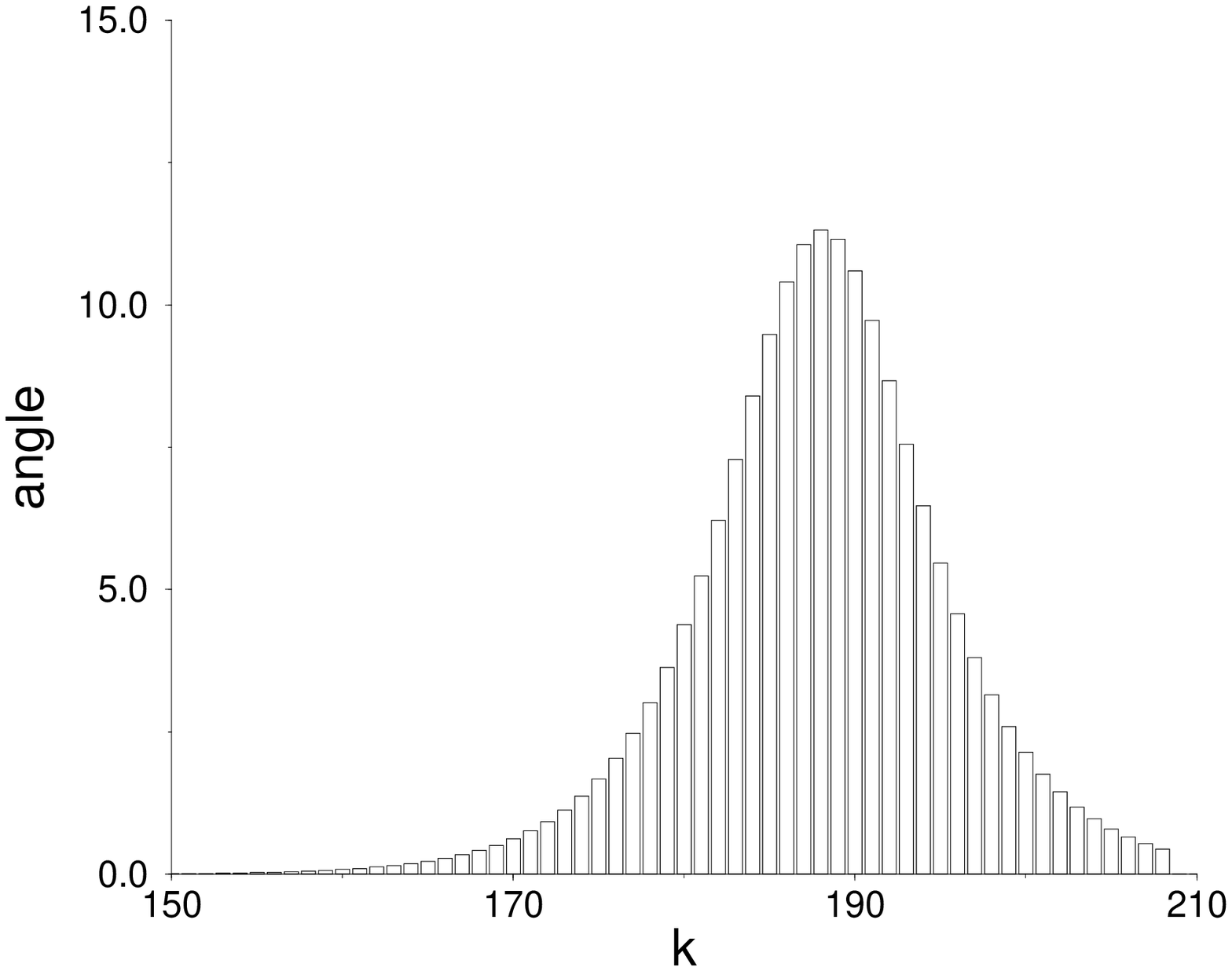,width=0.5\textwidth}}
\caption{Sequences of angles $\theta^k$ (omitting a long run-in and
tail of very small angles) derived from the procedure for making
near-optimal quantum Zeno protocols, with $\alpha_1=0.8$,
$\alpha_2=0.82$. The value of $(1-P(ident|1))(1-P(ident|2)) =
0.996621$ given by the algorithm is very close to $\eta^2=0.996620$.
Here $\epsilon$ was taken to be $0.0001$ and the value of $\lambda$
generated by the numerical search was 1.049996.}
\end{figure}

A second question concerns interaction-free discrimination of more
than two transparencies. What bounds apply in this case? We can also
broaden the question and consider situations where the object is not
destroyed when one particle interacts with it (eg the Elitzur-Vaidman
bomb), but where one wants to minimize the amount of interaction (eg
to reduce potential radiation damage). What bounds apply to {\em
minimal } absorption measurements \cite{KrennEtal00}?

%\subsection*{Acknowledgements}

We thank Richard Jozsa, Noah Linden, Sandu Popescu and Stefano Pironio
for helpful discussion. Particular thanks to Sandu Popescu for raising
the question of whether two grey levels can be distinguished by an
AFM.  S.M. is a research associate of the Belgian National Research
Fund. He thanks the European Science Foundation for financial support.

%\bibliographystyle{aaai.bst}
%\bibliography{gjm}

\end{multicols}

\end{document}